\documentclass[aps,prb,twocolumn]{revtex4}
\newcommand{\sign}{\mathop{\rm sign}\nolimits}
\renewcommand{\min}{\mathop{\rm min}\nolimits}
\renewcommand{\max}{\mathop{\rm max}\nolimits}
\begin{document}
\bibliographystyle{prsty}
\title{ \begin{flushleft}
{\small
\sc
PHYSICAL REVIEW B
\hfill
Volume, Number
\hfill
  1999 } \\
\vspace{2mm}
\end{flushleft}
 Auger-like Relaxation of Inter-Landau-Level  Magneto-Plasmon
Excitations in the Quantised Hall Regime
\vspace{-1mm}}
\author{
S. Dickmann${}\!\!$\renewcommand{\thefootnote}{\fnsymbol{footnote}}
\footnotemark[1]}
\address{
Institute for Solid State Physics of Russian Academy of Sciences,
142432 Chernogolovka, Moscow District, Russia}
\author{Y. Levinson}
\address{Department of Condensed Matter Physics, The Weizmann
Institute of Science,
76100 Rehovot, Israel\\
\smallskip
{\rm(Received June 1999)}
\bigskip\\
\parbox{14.2cm}
{\rm
Auger relaxation in 2D strongly correlated electron gas
can be represented as an
Auger-like process for neutral magnetoplasmon excitations.
The case of ``dielectric" state with lack of free electrons
(i.e. at integer filling $\nu$) is considered. Really the
Auger-like
process is a coalescence of two magnetoplasmons which are
converted into a single one of a different plasmon mode with
zero 2D wave-vector. This event turns out
to be energetically allowed for magnetoplasmons near
their roton minima where the spectrum has the infinite density
of states.  As a result the additional possibility appears
for indirect observation of the magnetorotons by means of
anti-Stokes Raman scattering.
We find the rate of this process employing the
technique of Excitonic Representation for the relevant matrix element
calculation.
}
}
\maketitle

Auger-type processes (APs) are believed to be the dominant
inter-Landau-level electron scattering mechanism when emission of
LO-phonons is suppressed off the magnetophonon resonance
conditions.
Auger scattering determines the population of Landau levels (LLs)
in cyclotron resonance \cite{Ma,Hs}, anti-Stokes hot
luminescence \cite{P}, and Integer Quantum Hall breakdown phenomena
\cite{ka95,ko96}. One-electron description of an
AP is scattering of two electrons at the same LLs resulting in
deexcitation of one of them to a lower LL and excitation of the
other to a higher LL. If this lower LL is partially filled in the
ground state of 2D electron gas (2DEG), then such a process
reduces the total number of excited electrons, providing the 2DEG
relaxation. This simple picture is based on LL equidistance and
seems to correspond to real situation such as in experiments at
$\nu< 1$ \cite{Ma}) or in the case of large LL numbers
of initially excited states \cite{ka95,ko96}.  On the other hand it
fails when Coulomb corrections to
energy of a free electron are significant and depend on LL number.
Moreover, near an integer $\nu$ the deficiency of unoccupied states
in the
almost filled LL leads to the conclusion that the usual AP relaxation
would become very rare as it would be a result of three-particle
collisions among two excited electrons and an effective
hole (unoccupied state at the LL filled in ground state).

It is meanwhile well known that strong Coulomb correlations in
the Quantum Hall regime renormalize drastically the 2DEG
excitation spectrum. The electron promoted from $n$-th LL to
$(n+m)$-th one and the effective hole left
at the $n$-th LL interact with each other; hence they should be
considered as a collective excitation. For integer filling the
spectrum, being of dielectric type (with Zeeman gap $|g\mu_bB|$ for
an odd $\nu$, and with cyclotron gap $\hbar\omega_c$ if $\nu$ is
even), is represented by chargeless excitations, namely: intra-LL
spin-waves ($m=0$), inter-LL cyclotron excitations without spin-flip
(so-called magnetoplasmons (MPs) with $m\ne 0$), and those with
spin-flip\cite{by81,ka84}. In this representation an Auger-type
process could be realized as a conversion of two MPs with energy
in the vicinity of $m\hbar\omega_c$ into one MP in the vicinity of
$2m\hbar\omega_c$.

The lowest energy MP with $m=1$ has pronounced roton type minimum
in the energy dependence $\epsilon (q)$ on the 2D wave vector ${\bf
q}$. \cite{le80,by81,ka84,ka85}. Near this
minimum the density of states is infinite and this is the reason
why the corresponding excitations, magnetorotons, were detected
by means of resonant combination back-scattering \cite{pi92,pi93,pi94}
though this detection is only possible due to breakdown of wave-vector
conservation (see discussion in Refs.\onlinecite{ms92,pi94}).
In the measured signal only one other peak
of the same MP mode just close to $\hbar\omega_c$
is observed. It corresponds to the MP with ${\bf q}$ near the
origin, and  satisfying the momentum conservation this peak
is more intensive even though the MPs at ${\bf q}=0$ have much
lower density of states \cite{ka85}. Important for a coalescence of
two MPs is the energetic possibility of their conversion into
some other excitation. We see that this process being allowed for
magnetorotons is forbidden for the MPs with ${\bf
q}=0$, since the energy of the final
is essentially higher than $2\hbar\omega_c$ due to
Coulomb corrections \cite{ka84}. Analogously
the coalescence is forbidden for two
MPs which are in the other ``suspicious" phase region, namely,
near the $\epsilon(q)$ maximum (not observed experimentally as yet)
where the density of states is also infinite. The energy of one
``two-cyclotron" MP is essentially lower than the combined
energy of such two MPs near the maxima. Thus, the mentioned
experimental
detection of magnetorotons and the energetic possibility of
the considered process are the reasons explaining our special
interest in the MPs coalescence near their roton minima. Moreover,
it is preferable to find out the generated ``two-cyclotron" MP in
the state with small 2D wave-vector, because in this case the
generated MP could be detected by anti-Stokes Raman scattering
like in the experiments of Refs. \cite{pi92,pi93,pi94}. This is why
we will present more detailed results exactly for this case. We
calculate the decay rate of such Auger-like process.

We solve this problem for the case of ``strong
magnetic field'', i.e. in the lowest-order approximation in the
small parameter $E_c/\hbar\omega_c$, where $E_{c}=e^2/\kappa l_{B}$
is a characteristic Coulomb energy for electron-electron ($e$-$e$)
interaction in 2DEG, $l_{B}$ being the magnetic length, and $\kappa$
the effective dielectric constant. (For $B=10\,$T$\quad
\hbar\omega_{c}=17.3\,$meV,
$l_{B}=8.1\,$nm and $E_{c}=14\,$meV).
It is well known that in this approximation the problem of
two-particle excitation spectrum for $\nu=integer$ can be solved
exactly \cite{by81,ka84}. Now, our task is the calculation
of the transition matrix element.

Further we employ the so-called
excitonic representation (ER), which is very advantageous
for excitations from a filled LL. Let us label a certain
one-electron state characterized by its LL and spin sublevel
by $a=(n_a,\sigma_a)$. Then the excitations may be considered as
effective excitons with energies
\begin{equation}
\label{d}
  \epsilon_{ab}(q)=
  \hbar\omega_cm+
  |g\mu_bB|\delta S_z + {\cal E}_{ab}(q),
\end{equation}
where $ m=n_b-n_a,\quad \delta S_z=\sigma_b-\sigma_a,\quad
(\sigma_a,\;\sigma_b=\pm1/2), $
and the energy ${\cal E}_{ab}$ has a Coulomb origin. It is
of the order of or smaller than $E_c$.

We restrict ourselves only to the case of $\nu=1$ considering
only  MPs with $n_{a}=0$ and $n_{b}=1,2$ and with
$\sigma_{a}=\sigma_{b}=+1/2$. In this case we change the
subscript $ab$ in Eq.(\ref{d}) to 01 or 02, respectively.
The analytical and numerical calculation of the excitation spectra
of these 01 and 02 MPs are presented in \cite{ka84} in the strict
2D limit (S2DL) when the thickness of the 2DEG $d$ satisfies the
condition $d\ll l_B$. In fact the spectra depend on $d$ but their
shape do not change qualitatively. The function ${\cal E}_{01}(q)$ has a
roton minimum at $q=q_0\approx 1.92/l_{B}$:
\begin{equation}
\label{min}
 {\cal E}_{01}(q)=\varepsilon_{0} + (q-q_0)^2/2M,
  \quad     |q-q_0|\ll q_0\,,
\end{equation}
where in S2DL $M^{-1}\approx 0.28\,E_{c}l_{B}^2$ and $\varepsilon_0
\approx 0.15\,E_{c}$. The dependence ${\cal E}_{02}(q)$ is also
nonmonotonic, but in the range $0<ql_B<2.5$ does not change more than
$0.07\,E_c$. Of special importance is the difference $\delta={\cal
E}_{02}(0)-2\varepsilon_{0}$, which ``casually" is numerically small
in the scale
of $E_c$, namely in S2DL $\delta \approx 0.019\,E_{c}\simeq 3\div 4\,$
K for $B=10\div 20\,$T, but is positive \cite{fag6}. The desired matrix
element of the considered conversion is
\begin{equation}
\label{m}
{\cal M}({\bf q}_{1},{\bf q}_{2})
  ={}_{02}\langle {\bf q}_{1}+{\bf q}_{2};1|
  H|{\bf q}_{1},{\bf q}_{2};2 \rangle_{01} .
\end{equation}
Here $H$ is Hamiltonian, the initial state is a two 01MP state,
and the final one is a one 02MP state.

The total 2DEG Hamiltonian is $H=H_0+H_{int}$,
where the Hamiltonian of the noninteracting electrons is
\begin{equation}
\label{Ho}
H_{0}=\sum_{n,p,\sigma}\left[(n+1/2)\hbar\omega_{c}-
  |g\mu_{b}B|{\sigma}\right]e^+_{n,p,\sigma}
  e_{n,p,\sigma}\,.
\end{equation}
Here $e_{n,p,\sigma}$ is the electron annihilation operator
at $n$-th LL having $\sigma$ as the ${\hat z}$-component of spin,
$p=k_y$ is the intra-level Landau gauge quantum number.
Within the framework of strong magnetic field approximation it
is enough to keep in the interaction Hamiltonian,
$H_{int}$, only
the terms which conserve cyclotron part of the energy,
or in other words, the terms which commute with $H_{0}$.
The Coulomb part of the Hamiltonian may therefore be written
in the form
\begin{eqnarray}
\label{Hint}
\nonumber{H_{int}={\cal N}^{-1}
  \sum_{p,p',{\bf q}\atop n,m,l,k,\sigma_1,\sigma_2}
  V_{nmlk}({q})\exp{[iq_{x}(p'-p)]}\;\cdot}\\
  {\displaystyle e^+_{n,p+q_y,\sigma_1}e^+_{m,p',\sigma_2}
  e_{l,p'+q_y,\sigma_2}e_{k,p,\sigma_1},}
\end{eqnarray}
which provides automatically the cyclotron energy
conservation rule $n+m=l+k$, because
\begin{equation}
\label{V}
V_{nmlk}({q})= (2\pi)^{-1} V(q) h_{nk}({\bf q})
  h_{lm}^*({\bf q})\delta_{n+m,l+k}.
\end{equation}
We use now dimensionless length and wave-vectors measured in the
units of $l_{B}$ and $l_{B}^{-1}$.
${\cal N}=L^2/2\pi l_B^2$ is the total number of
magnetic flux quanta in the normalization area $L^2$,
and $V(q)$ is the 2D Fourier component of the Coulomb potential
averaged with the wave function in the ${\hat z}$ direction
(so that in S2DL: $V(q)=2\pi E_c/q$), and
\begin{eqnarray*}
\label{h}
h_{nk}({\bf q})=
{\displaystyle
\int_{-\infty}^{+\infty}}\!\!dx
  \chi_n(x+q_y/2) e^{iq_xx}
  \chi_k(x-q_y/2)={}\qquad{} \\
  \left[\frac{\displaystyle\min(n,k)!}
  {\displaystyle\max(n,k)!}\right]^{1/2}
    \left[\frac{\displaystyle iq_x +q_y\sign{(n-k)}}
  {\displaystyle \sqrt{2}}\right]^{|n-k|}\cdot\\
 e^{-q^2/4} {\displaystyle L_{\min(n,k)}^{|n-k|}}
         (q^2/2)
\end{eqnarray*}
$\chi_n(x)$ is the normalized $n$-th harmonic oscillator
function, $L_{n}^{j}$ is Laguerre polynomial.

Now we define in ER the states in the matrix element (\ref{m})
in order to calculate the last one. Let $a$ be the filled LL,
i.e. in our particular case $a=(0,1/2)$. We designate
$a_{p}\equiv e_{n_{a},p,\sigma_{a}}$ while
 $b_{p}\equiv e_{n_{b},p,\sigma_{b}}$
for every other one-electron state $b$.
The ER means a
replacement of operators $e^+_{n,p,\sigma}$ and $e_{n,p,\sigma}$
by a set of inter-LL
``excitonic" creation and annihilation operators for $a\not=b$
(i.e. $n_b\not=n_a$, or $\sigma_a\not=\sigma_b$)
\begin{equation}
\label{Q}
{}\!{\cal Q}_{ab{\bf q}}^{+}=\frac{1}{\sqrt{\cal N}}\sum_{p}\,
e^{-iq_x p}
  b_{p+\frac{q_y}{2}}^{+}\,a_{p-\frac{q_y}{2}},\;\,
  {\cal Q}_{ab{\bf q}}={\cal Q}_{ba\,{-{\bf q}}}^{+},\!
\end{equation}
and intra-LL ``displacement'' operators $A_{\bf q}$
and $B_{\bf q}$ (see Ref. \cite{di96}). We do not write here the latter
ones, because
they, being required for total ER of Hamiltonian (\ref{Hint}), are
not used directly for matrix element (\ref{m}) calculation.

Some commutation rules for operators
(\ref{Q}) are the same as the ones
obtained in Ref. \cite{di96} minding the case $a=(n,1/2)$,
$b=(n,-1/2)$. We derive the additional ones considering $a\not=b\not=c$:
\begin{eqnarray}
\label{com}
\nonumber{
[{\cal Q}_{bc{\bf q_1}}^{+},
  {\cal Q}_{ab{\bf q_2}}]=0,
\;\,[{\cal Q}_{bc{\bf q_1}}^+,{\cal Q}_{ab{\bf q_2}}^+]
}\\
  =\frac{e^{-i\Theta_{12}}}{{\cal N}^{1/2}}
   {\cal Q}_{ac\,{\bf q_1+q_2}}^+.
\end{eqnarray}
Here
  $\Theta_{12}=\Theta({\bf q}_{1},{\bf q}_{2})=
  ({\bf q}_{1}\times {\bf q}_{2})_{z}/2=
  \frac{1}{2} q_{1}q_{2}\sin\alpha$,
 where $\alpha$ is an angle between ${\bf q}_{2}$ and ${\bf q}_{1}$.
Note that the considered operators were employed earlier in some other
form as applied to ``valley-wave"
excitations\cite{ra86} and, also, to spin-waves\cite{di96,by96},
when $m=0$, $|\delta S_z|=1$.

The operator ${\cal Q}_{ab{\bf q}}^+$ creates a $ab$MP:
$|{\bf q};1\rangle_{ab}={\cal Q}_{ab{\bf q}}^{+}|0\rangle$.
Here $|0\rangle$ is the ground state where
the level $a$ is fully occupied,
whereas $b$ is empty: $a^{+}_{p}|0\rangle=b_{p}|0\rangle\equiv 0$.
This is equivalent to identities
$
  A^+_{\bf q}|0\rangle\equiv
                \delta_{0,{\bf q}}|0\rangle\quad \mbox{and}\quad
  B^+_{\bf q}|0\rangle\equiv{\cal Q}_{ab{\bf q}}|0\rangle\equiv 0\,.
$

The choice of the state $b$ depends on
a type of problem. In our case the states entering the matrix element
(\ref{m}) are
\begin{equation}
\label{s}
|{\bf q};1\rangle _{02}= {\cal Q}_{02\,{\bf q}}^{+}|0\rangle,\quad
|{\bf q}_1,{\bf q}_2;2 \rangle _{01}= {\cal Q}_{01{\bf q}_1}^{+}
  {\cal Q}_{01{\bf q}_2}^{+}|0\rangle,
\end{equation}
As above 01 and 02 stand for $ab$ with $a=(0,1/2)$ and $b=(1,1/2)$
or $b=(2,1/2)$ respectively. These states are orthogonal and are
eigenstates of the Hamiltonian $H$ in the limit ${\cal
N}\rightarrow\infty$, i.e.
$$
\begin{array}{l}
H_{int}|{\bf q};1\rangle_{02}=[E_{0}+{\cal E}_{02}(q)]|{\bf
q};1\rangle_{02}+\cdots,\\
H_{int}|{\bf q}_1,{\bf q}_2;2\rangle_{01}=\\
\nonumber{
\qquad{}\qquad{}\quad{} [E_{0}+{\cal E}_{01}(q_{1})+{\cal E}_{01}(q_{2})]|{\bf q}_1,
 {\bf q}_2;2 \rangle_{01}+\cdots}\;\;,
\end{array}
$$
where $E_0$ is the Coulomb ground state energy ($H_{int}|0\rangle=
E_0|0\rangle$) and the dots
correspond to some states having a norm of the order of $E_c/{\cal
N}$. The states (\ref{s}) are the correct initial and final
states in the scattering problem for a low density gas of MPs, but
the scattering matrix element (\ref{m}) has to be calculated with
higher accuracy, since ${\cal M}\sim {\cal N}^{-1/2}$.

Instead of the value (\ref{m}) it is more convenient to calculate the
conjugate one ${\cal M}^*$ substituting in Eq. (\ref{m}) the
expressions (\ref{s}). After one has done the ER transformation of the
Hamiltonian (\ref{Hint}) in terms of operators (\ref{Q}) together with
$A_{\bf q}$ and $B_{\bf q}$.
then taking into account the properties of the ground state
$|0\rangle $ and the commutation rules (\ref{com}) one finds that the
only term of Hamiltonian which contributes to the matrix element
(\ref{m}) is
$$\sum_{\bf q} {V}_{1120}(q)
  {\cal Q}^+_{01{\bf q}}{\cal Q}_{12{\bf q}}\,.
$$
Using again as tools the properties of operators
(\ref{Q}) and of the state $|0\rangle$ we obtain for
${\bf q}_{1}\not={\bf q}_{2}$
\begin{eqnarray}
\label{mm}
\nonumber{{\cal M}({\bf q}_{1},{\bf q}_{2})=
{\cal N}^{-1/2}\left[
  u(q_1) e^{-i\Theta_{12}}+u(q_2) e^{i\Theta_{12}}\right.}\\
   -\left.v(q_1) e^{-i\Theta_{12}}-
  v(q_2) e^{i\Theta_{12}}\right],
\end{eqnarray}
where
\begin{eqnarray*}
\label{}
   u(q)&=&(2^{5/2}\pi)^{-1}q^2V(q)[2-(ql_{B})^{2}/2]
e^{-(ql_{B})^{2}/2},\\
   v(q)&=&l_B^2\int_0^{\infty}dp p u(p)J_0(pql_{B}^{2})\,.
\end{eqnarray*}
We returned here to dimensional quantities (also
suitable
redefinition is $V(q)\to l_B^2V(q)$). In S2DL one can find
analytic expression of  $v(q)$, involving polynomials,
exponentials, and modified Bessel functions.

The depopulation rate of a 01MPs due to
their coalescence is
\begin{eqnarray}
\label{mer}
 \nonumber {\cal R}={1\over 2}\sum_{{\bf q}_1,{\bf q}_2}\frac{2\pi}{\hbar}
  \left|{\cal M}({\bf q}_1,{\bf q}_2)\right|^2
  \overline{n}({\bf q}_1) \overline{n}({\bf q}_2)
  \cdot\\
  \delta\left[{\cal E}_{01}({\bf q}_1)+{\cal E}_{01}({\bf q}_2)
  -{\cal E}_{02}({\bf q}_1+{\bf q}_2)\right]
\end{eqnarray}
where $\overline{n}({\bf q})$ are occupation numbers of
01MPs. We consider the occupancy for 02MPs to be
small and do not take into account
corresponding stimulated processes and 02MP decay.
(Note that the terms with ${\bf q}_1={\bf q}_2$ give no
essential contribution to this sum).

Because of the mentioned reasons, we will consider a special
situation
when the 01MPs occupy states near the magnetoroton minima
and calculate the rate of depopulation due to creation
of 02MPs only with small $q$. These 02MPs can be detected
by anti-Stokes Raman scattering like in experiments
\cite{pi92,pi93,pi94}.
For this purpose we have to sum in Eq.(\ref{mer}) with
the restriction $|{\bf q}_{1}+{\bf q}_{2}|<{\tilde q}$
and we will show later that ${\tilde q}\ll l_{B}^{-1}$.
Under these assumptions we can put in $|{\cal M}|^2$
${\bf q}_{1}=-{\bf q}_{2}$ and $|{\bf q}_{1}|=|{\bf
q}_{2}|=q_{0}$.
We also assume that $\overline{n}({\bf q})=\overline{n}$
can be considered to be constant as long as due
to energy conservation we are dealing with
the a narrow band determined by inequalities: $\varepsilon_0<
{\cal E}_{01}(q)<{\cal E}_{02}({\tilde q})-\varepsilon_0$.
Using this simplifications and replacing summation by
integration $\sum_{{\bf q}}={\cal N}l_{B}^{2}\int d^{2}q/
(2\pi)$
we find the rate of 02MP creation with $|{\bf q}|<{\tilde q}$
per unit area to be
\begin{equation}
\label{r}
\frac{{\cal R}(|{\bf q}|<{\tilde q})}{L^{2}}=
\displaystyle\frac{\overline{n}^{2}l_B^2q_0}{2\hbar}
[u(q_0)-v(q_0)]^2 \left({M\over\delta}\right)^{1/2} {\tilde q}^{2}.
\end{equation}

The question to be considered is the role of the random impurity
potential $U({\bf r})$ which was neglected in the above
calculations. The distance between an excited electron and a hole
in real space is $l_B^2{\bf q}\times {\hat z}$ (see
Refs.\cite{le80,by81,ka84}). Assuming $U({\bf r})$ to be
smooth (correlation length $\Lambda\gg l_{B}$) one can find that
the energy correction for a MP with the wave vector ${\bf q}$.
In the dipole approximation it is
$\delta {\cal E}({\bf q},{\bf r})=-\hbar {\bf q}{\bf v}_d$
for any $ab$MP, where ${\bf v}_d=({\hat z}\times\nabla U({\bf r}))
l_B^2/\hbar$ is the drift velocity. This additional energy leads to
inhomogeneous broadening of the MP energy.
One can see that the random potential correction plays no significant role if
\begin{equation}
\label{dcr}
|d{\cal E}_{ab}/dq|\gg l_B^2\left|{\bf \nabla}U\right|
\end{equation}
which means that the  electron-hole Coulomb
interaction is stronger than the force the electron and the effective
hole
are subjected to in the random electric potential. Evidently the
other meaning of this condition is that the exciton velocity has to
be greater
than the drift velocity in the external field \cite{fag6}.
Alternatively, we have two independent
quasiparticles, electron and hole, whose motion is determined
mainly by the random potential
and the $e$-$e$ interaction has to be considered only as a
perturbation \cite{le95}.
For $q\simeq l_{B}^{-1}$ one can
estimate the inhomogeneous broadening
$\delta{\cal E}\simeq \Delta (l_{B}/\Lambda)$,
where $\Delta$ is the random potential amplitude (i.e.
$\nabla U\sim \Delta/\Lambda$).
With typical values $\Lambda=50$nm and  $\Delta=1$meV
one finds $\delta{\cal E}=0.2$meV.
This is small compared to the width of the MP band
and small compared to $\epsilon_{0}$
but of the same order as the energy $\delta$
relevant for energy conservation. At the same time,
since in the dipole approximation the inhomogeneous broadening
of the level ${\cal E}_{ab}({\bf q})$ do not depend on $ab$
it gives no contribution to the delta-function argument in
Eq.(\ref{mer}). Higher order corrections to $\delta{\cal E}$ are of
the order of $\Delta (l_{B}/\Lambda)^{2}\simeq 0.04\,$meV and small
compared even to $\delta$.
As a result we conclude that the role of the random potential
is negligible compared to the $e$-$e$ interactions.

However the role of the random potential is crucial in determining
the cutoff ${\tilde q}$. The momentum of a 02MP detected by
anti-Stokes Raman scattering is defined from momentum conservation
as ${\bf q}={\bf k}_{2\parallel}-{\bf k}_{1\parallel}$, where
${\bf k}_{1\parallel}$ and
${\bf k}_{2\parallel}$ are the ``in-plane" wave vector components of the
incident and scattered photons. In the case of no disorder the cutoff
${\tilde q}$ is defined by the uncertainty of
${\bf k}_{2\parallel}-{\bf
k}_{1\parallel}$, i.e. by the
spectral resolution and the geometry of the optical
experiment. This uncertainty is $<10^4\,$cm$^{-1}$ according
to Refs. \cite{pi92,pi93,pi94} and
the cutoff ${\tilde q}$ actually comes from the disorder which
violates momentum conservation.

In the approximation of S2DL one may estimate
$d{\cal E}_{02}/dq\simeq  E_cq^2l_B^3$ for $ql_{B}\ll 1$
\cite{fag5},
and the uncertainty  of $q$ due to disorder
can be found from Eq.(\ref{dcr}) giving
${\tilde q}\simeq (\Delta/E_{c})^{1/2}(\Lambda
l_{B})^{-1/2}$.
This value does not depend on the magnetic field and
for the used numerical parameters ${\tilde q}\sim
10^5\,$cm${}^{-1}$. The substitution into Eq.(\ref{r})
gives
\begin{equation}
\label{r2}
{\cal R}(|{\bf q}|<{\tilde q})/L^{2}
\sim 0.05\cdot\frac{\overline{n}^2\Delta}{\hbar l_B\Lambda}
\;\;\propto\;\; B^{1/2}
\end{equation}
(it is taken into account that $u(q_0)-v(q_0)\approx
-0.062E_C$).

Now let us estimate the total decay of 01MPs supposing that the most of
them are concentrated in the vicinity of the roton minima.
Generally, a more complicated summation (\ref{r}) has to
be fulfilled in this case, because the allowed phase region where
02MPs can be generated is not small. Indeed, the very weak
dependence ${\cal E}_{02}(q)$ in its initial spectrum portion
leads to the only condition $q\lesssim l_B^{-1}$ for allowed O2MP
wave-vectors. However to obtain the approximate total rate of the
coalescing 01MPs the formula (\ref{r2}) can be exploited again.
Estimating the 01MP density near their roton minima as $N\simeq
\overline{n}q_0(2M\delta{\cal E})^{1/2}$ (because the roton minima
broadening due to inhomogeneity is $|{\bf q}-{\bf q}_0|\sim
(2M\delta{\cal E})^{1/2}$) and setting $dN/dt$
equal to decay rate (\ref{r2}) with ${\tilde q}\sim l_B^{-1}$ we find
the characteristic relaxation time
$\tau=\overline{n}dt/d\overline{n}$ which turns out to be
$$
  \tau \sim 10^2\hbar(\Delta l_B/E_C^3\Lambda)^{1/2}/\overline{n}\sim
  1/\overline{n}\,\mbox{ps}
$$
(therefore $\tau \propto 1/B$). This value
should be for real experiments compared with time characteristics of
other possible relaxation channels, for example when the conditions
of magnetophonon resonance are satisfied.

The value $\overline{n}$ remains indefinite because it depends on the
specific manner of 01MPs excitation. We think the photoluminescence
excitation technique is likely to be more appropriate for it,
as far as therewith the excitation would occur in two independent
steps: namely, by generation of an electron at the 1-th LL and a
hole in the valence band, and by recombination of some electron
from the filled LL with the hole. As a result 01MPs
with various ${\bf q}$-s can appear. This technique should be more
effective for magnetoroton excitation in comparison with Refs.
\cite{pi92,pi93,pi94} though in itself it does not permit to
detect the magntorotons. Nevertheless, if one simultaneously could
find 02MPs by means of anti-Stokes Raman scattering or by means of
hot luminescence from the 2-nd LL, it would be an indirect
confirmation of the presence of 01MPs near their roton minima. Note also
that the appropriate consideration of kinetic relations shows that
the occupation number for 02MPs could be
expected to be of the order of $\overline{n}^2$ once the
quasi-equilibrium $2\times$01MP$\leftrightarrow$02MP is
established.

S.D. thanks for hospitality the Department of Condensed Matter
Physics of Weizmann Institute of Science, where the main
part of this work was done. The work is supported by the MINERVA
Foundation and by the Russian Fund for Basic Research
(Project 99-02-17476).

\renewcommand{\thefootnote}{\fnsymbol{footnote}}
\footnotetext[1]{Electronic address: dickmann@issp.ac.ru}


\begin{references}
\bibitem{fag5}
More accurately the analytical expression for ${\cal E}_{02}(q)/E_c$
(see Ref. [8]) contains the lowest order term $c(ql_q)^2$ near $q=0$,
but the factor $c$ is in S2DL numerically small.
\bibitem{fag6}
If $\delta$ turns out to be negative, then the direct transition
(i.e. without acoustic phonon participation) would be possible
only for generation 02MP with rather large wave-vector, which is
outside of region required for detection (see text).
\end{references}
\end{document}